\documentclass[aps, prb,amsmath,amssymb,onecolumn,showpacs]{revtex4-2}
\usepackage{graphicx}
\usepackage{physics}
\usepackage{caption}
\usepackage{subcaption}

\def\Xint#1{\mathchoice
   {\XXint\displaystyle\textstyle{#1}}%
   {\XXint\textstyle\scriptstyle{#1}}%
   {\XXint\scriptstyle\scriptscriptstyle{#1}}%
   {\XXint\scriptscriptstyle\scriptscriptstyle{#1}}%
   \!\int}
\def\XXint#1#2#3{{\setbox0=\hbox{$#1{#2#3}{\int}$}
     \vcenter{\hbox{$#2#3$}}\kern-.5\wd0}}

\def\dashint{\Xint-}

\begin{document}
\title{Coexistence of localized and extended states in the Anderson model with long-range hopping}

\author{V. Temkin}
\email{vatemkin@edu.hse.ru}
\affiliation{Condensed-matter physics laboratory, National Research University Higher School of Economics, Moscow 101000, Russia,}
\author{A. S. Ioselevich}
\email{iossel@itp.ac.ru}
\affiliation{Condensed-matter physics laboratory, National Research University Higher School of Economics, Moscow 101000, Russia,}
\affiliation{L. D. Landau Institute for Theoretical Physics, Moscow 119334, Russia}

\date{\today}
 
\begin{abstract}
We study states arising from fluctuations in the disorder potential in systems with long-range hopping. Here, contrary to systems with short-range hopping, the optimal fluctuations of disorder responsible for the formation of the states in the gap, are not rendered shallow and long-range when $E$ approaches the band edge ($E\to 0$). Instead, they remain deep and short-range. The corresponding electronic wave functions also remain short-range-localized for all $E<0$. This behavior has striking implications for the structure of the wave functions slightly above $E=0$. By a study of finite systems, we demonstrate that the wave functions $\Psi_E$ transform from a localized to a quasi-localized type upon crossing the $E=0$ level, forming resonances embedded in the $E>0$ continuum. The quasi-localized $\Psi_{E>0}$ consists of a short-range core that is essentially the same as $\Psi_{E=0}$ and a delocalized tail extending to the boundaries of the system. The amplitude of the tail is small, but it decreases with $r$ slowly. Its contribution to the norm of the wave function dominates for sufficiently large system sizes, $L\gg L_c(E)$; such states behave as delocalized ones. In contrast, in small systems, $L\ll L_c(E)$, quasi-localized states are overwhelmingly dominated by the localized cores and are effectively localized.
\end{abstract}

\maketitle


\section{Introduction }

The theoretical and numerical study of eigenfunctions for the quantum-mechanical problem with deterministic power-law hopping
\begin{eqnarray}
\hat{H}_{\rm hop}=\sum_{\bf jj^\prime}\varepsilon_{\bf j-j^\prime}a^\dagger_{\bf j}a_{\bf j^\prime},\quad \varepsilon_{\bf r} \propto r^{-\beta},\label{h0}
\end{eqnarray}
and local disorder 
\begin{eqnarray}
\hat{H}_{\rm dis}=\sum_{\bf j}V_{\bf j}a^\dagger_{\bf j}a_{\bf j},\label{h01}
\end{eqnarray}
which is a modification of the Anderson impurity model~\cite{AndersonLoc}, first started more than 30 years ago \cite{Burin89} (or see~\cite{Levitov89, Levitov90} which are closely related) and has attracted significant interest in the community \cite{malyshev2004monitoring, deng2018duality, deng2016quantum,deng2022superdiffusion,tang2022nonergodic, klinger2021single, cantin2018effect, rodriguez2000quantum, rodriguez2003anderson, nosov2019correlation, nosov2019robustness, syzranov2015critical, syzranov2016multifractality, Syz15, Syz-review, mirlin2000statistics, wegner1980inverse}. Today, the demand for proper theoretical analysis is great because of the growing number of experimentally accessible physical systems that are described by the same mathematical framework. For example, it can be used to describe quantum superconductor-metal transition in 2D disordered metals \cite{TF2020} or the behavior of arrays of trapped ions \cite{1DAtoms, 2DAtoms}, which is of great interest in quantum computing (for more examples, see \cite{Syz-review}).

In this study, we consider the case when the value of the exponent $\beta$ in \eqref{h0} lies in an interval $D < \beta < 3D/2$, where $D$ is the dimension of the considered lattice (we provide analytical results for any dimension, but our numerical study of the optimal fluctuation, see Section~\ref{Numerical study of the optimal fluctuation}, is limited to physical dimensions $D = 1,2,3$ only). In the case that we explore, the effects of typical weak fluctuations of the random potential were studied extensively, and it was shown \cite{Syz15} that a non-Anderson disorder-driven metal-insulator transition takes place. Here, we aim to elaborate on the understanding of the effects of the interplay between typical weak fluctuations of the random potential and rare strong local fluctuations, (the latter are sometimes called ``rare-regions''). Particularly, we explain numerical results from \cite{TikhonovIoselevichFeigel'man}, that seem to indicate the possibility of the coexistence of localized and extended states near one of the edges of the band in the considered model. We expect the effects of strong local fluctuations to be the main mechanism for the formation of small-sized localized (or rather quasi-localized, see below) states on the background of extended ones. As will become clear later in this paper, no ``true'' coexistence is present in the investigated case, and Mott's principle \cite{mott1967electrons} is not violated.


The localized band-gap states arising due to localized fluctuations in a standard Anderson model with nearest neighbor hopping and gaussian disorder in dimensions $D\leq 3$ are well known -- they form the so-called Lifshitz tail in the density of states $\nu(E)$ within the energy gap (see \cite{Lifshitz}). For $E$  being deep enough in the gap, $\nu(E)$ is exponentially small
\begin{eqnarray}
\nu(E)\propto\exp\{-S_{\rm Lif}(E)/W^2\},\quad S_{\rm Lif}(E)\propto |E|^{2-D/2}
\label{lii1}
\end{eqnarray}
where $W^2=\langle V^2\rangle$.
Here the energy $E$ is accounted for with respect to the band edge.
The optimal fluctuation of disorder, responsible for formation of the localized state with energy $E$ has  a spatial scale $a(E)$ and depth $U(E)$, where 
\begin{eqnarray}
 a(E)\propto |E|^{-1/2}, \quad U(E)\sim |E|.
\label{lii3a}
\end{eqnarray}
The optimal fluctuation approach (that is, technically, the steepest descent method for the functional integration over configurations of random potential) is justified if $S_{\rm Lif}(E)/W^2\gg 1$

Note that $S_{\rm Lif}(E)\to 0$ and $a(E)\to \infty$ as $E\to 0$, so that the optimal fluctuation method is not applicable in the close vicinity if the band edge.

Generalization of the result \eqref{lii1} to the systems with the general hopping Hamiltonian \eqref{h0} gives
\begin{eqnarray}
 S_{\rm Lif}\propto |E|^{2-D/\alpha},\quad \alpha\equiv\beta- D.
\label{lii3}
\end{eqnarray}
The result \eqref{lii3} is perfectly reasonable for $2-D/\alpha>0$, the estimates \eqref{lii3a} apply to the optimal fluctuation in this case.  The situation is changed cardinally  for $2-D/\alpha<0$: here the fluctuation with size $a(E)\propto |E|^{-1/2}$ ceases to be an optimal one: the actual ''non-Lifshitz'' optimal fluctuation at $2-D/\alpha<0$ has a microscopic spatial scale $a_0$:
\begin{eqnarray}
 S_{\rm nonLif}(E)\approx S_{\rm nonLif}(0)+A|E|,\quad a(E)\sim a_0,\label{liip0}\\
 S_{\rm nonLif}(0)\sim \varepsilon_0^2,\quad A\sim 1/\varepsilon_0
\label{liip}
\end{eqnarray}
where $\varepsilon_0$ is some characteristic energy scale of order of the electronic bandwidth. The linear expansion \eqref{liip0} is valid for $|E|\ll \varepsilon_0$

 It is important, that, in contrast with the long-range Lifshitz fluctuations, the short-range non-Lifshitz optimal fluctuations provide a valid description of the corresponding contribution to the density of states  even at $E\to 0$: the corresponding $S_{\rm nonLif}(E)$ tends to a finite limit as $E\to 0$. The latter observation was the origin for the idea about the existence of Lifshitz-like states not only for $E<0$, but also for  $E>0$, at least in a certain range.
 
 To reliably address the question of possible existence of localized electronic states  on the continuum background of the delocalized band states, one is forced to consider finite systems.  As we will see, the structure of both delocalized and quasi-localized states essentially depends on the system size. Namely, we show that upon crossing the band edge $E=0$ the true localized states that existed for $E<0$, continually transform into the quasi-localized ones. They consist of the localized parts (which are basically are the same as for $E<0$) and the delocalized ones with the amplitude that vanish continuously as $E$ approaches 0 from above. The delocalized part is, however, extremely sensitive to the systems size $L$ and becomes increasingly important with increasing $L$. As a result, the quasi-localized states behave practically as localized ones for $L<L_c(E)\equiv E^{-\frac{D+1}{\alpha}+2}$ while becoming essentially delocalized for $L>L_c(E)$.

\section{The problem statement. }

We consider a FINITE $D$-dimensional hypercubic lattice of $(2L)^D$ sites ($L\gg 1$) with periodic boundary conditions. The hamiltonian
\begin{eqnarray}
\hat{H}=\hat{H}_{\rm hop}+\hat{H}_{\rm dis},\label{h1}
\end{eqnarray}
where the random potential $V_{\bf j}$ obeys the gaussian distribution: 
\begin{eqnarray}
{\cal P}\{V\}=\prod_{\bf j}P(V_{\bf j})\propto e^{-\frac{S\{V\}}{W^2}},\quad S\{V\}=\frac{1}{2}\sum_{\bf j}V^2_{\bf j},\nonumber\\ 
P(V)=\frac{1}{\sqrt{2\pi}W}\exp\left\{-\frac{V^2}{2W^2}\right\}.
\label{h1er}
\end{eqnarray}

In the momentum representation
\begin{eqnarray}
\hat{H}=\sum_{n}\varepsilon({\bf k_n})a^\dagger_{\bf n}a_{\bf n}+\sum_{\bf nn^\prime}a^\dagger_{\bf n}a_{\bf n^\prime}V_{\bf n^\prime-n},\label{h2}
\end{eqnarray}
where the momenta ${\bf k_n}\equiv \pi {\bf n}/L$, and the corresponding normalized eigenfunctions
\begin{eqnarray}
\phi_{\bf n}({\bf j})=(2L)^{-D/2}\exp(i \pi {\bf (j \cdot n)}/L),\quad {\bf n}\equiv (n_1,n_2,\ldots n_D),\quad n_i=-L,-L+1,\ldots, L,
\label{h3}
\end{eqnarray}
\begin{eqnarray}
a_{\bf j}=\sum_{\bf n}a_{\bf n}\phi_{\bf n}({\bf j}),\quad a^\dagger_{\bf j}=\sum_{\bf n}a^\dagger_{\bf n}\phi^*_{\bf n}({\bf j}),
\end{eqnarray}
The kinetic energy  in  $k$-representation:
\begin{eqnarray}
\varepsilon({\bf k})=\sum_{\bf jj^\prime}\varepsilon_{\bf j-j^\prime}\phi_{\bf n+k}({\bf j})\phi_{\bf n}({\bf j^\prime})=\varepsilon_0f({\bf k}),\quad {\bf k}=\frac{\pi {\bf n}}{L},\quad {\bf k}\equiv (k_1,k_2,\ldots k_D)\quad -\pi<k_i<\pi,\label{h4}
\end{eqnarray}
where all lengths are measured in the units of lattice spacing. The characteristic energy $\varepsilon_0$ by the order of magnitude is an electronic bandwidth, in what follows we will measure all energies in the units of $\varepsilon_0$. The $2\pi$-periodic function $f(k)$ 
\begin{eqnarray}
f( {\bf k})=\left|4\sum_{\mu=1}^D\sin^2 k_\mu/2\right|^{\alpha/2},\quad f_{\max}=f(\pi,\pi,\ldots,\pi)=(4D)^{\alpha/2},\quad \alpha=\beta- D.\label{h5}
\end{eqnarray}
behaves at $k\ll 1$ as 
\begin{eqnarray}
f({\bf k})\approx |k|^{\alpha},\quad |k|\equiv\left(\sum_{\mu=1}^D k_\mu^2\right)^{1/2}.\label{h6}
\end{eqnarray}
Thus, all the energies are confined within the interval $0<\varepsilon_n<W_{\rm band}$, where $W_{\rm band}=\varepsilon_0f_{\max}$.

\section{Low energy properties of an ideal system }

For small $E\ll 1$ the spectrum $\varepsilon({\bf k})$ is isotropic and the corresponding wave-functions can be characterized by the angular momenta. In our problem only the fully symmetric solutions are relevant, because the low-symmetric ones vanish at $r\to 0$ and hardly feel the strongly localized potential $V_{\bf j}$. The normalized fully-symmetric eigenfunctions are
\begin{eqnarray}
\psi_n(r)=\sqrt{\frac{2k_n^{D-1}}{\sigma_D L}}f(k_nr),\quad \int_0^{L} \sigma_D r^{D-1}dr|\psi_n(r)|^2=1,\nonumber\\
f(x)=\sqrt{\frac{\pi}{2}}\frac{J_{D/2-1}(x)}{x^{D/2-1}},
\quad \sigma_D=\frac{D\pi^{D/2}}{\Gamma{\left(D/2 + 1\right)}}
\end{eqnarray}
where $r\equiv |{\bf r}|$, $k\equiv |{\bf k}|$, $k_n=\pi n/L$, $\sigma_D$ being the surface area of $D$-dimensional sphere with unit radius. The asymptotics of $f(x)$ are
\begin{eqnarray}
f(x\gg1)\approx x^{-\frac{D-1}{2}}
\cos (x+\varphi_D), \quad
\varphi_D=\frac{\pi}{4}(1-D),\quad f(0)=\frac{\sqrt{\pi/2}}{2^{D/2-1}\Gamma(D/2)}.
\end{eqnarray}
For general $D$ the low energy ($E\ll 1$) density of states
\begin{eqnarray}
\nu_0^{(D)}(E)=\frac{\sigma_D}{(2\pi)^D}\frac{k^{D-1}dk}{dE}=\frac{\sigma_D/D}{(2\pi)^D}\frac{d(k^{D})}{dE}=\frac{\sigma_D/D}{(2\pi)^D}\frac{d(E^{D/\alpha})}{dE}=\frac{\sigma_D}{(2\pi)^D\alpha}\frac{K^D}{E}.
\label{denst1}
\end{eqnarray}
We have introduced characteristic momentum
\begin{eqnarray}
K=E^{1/\alpha},
\end{eqnarray}
which, alongside with the short range scale $K_0\sim\pi$ is an  important momentum scale in our problem. Throughout this paper we assume that
\begin{eqnarray}
1/L\ll K\ll 1
\label{highExc}
\end{eqnarray}

The general level spacing, which takes into account all the states irrespective to their symmetry, is 
\begin{eqnarray}
\delta_D(E)=\left(\nu_0^{(D)}(E)L^D\right)^{-1}=\frac{\left(2\pi\right)^D\alpha}{\sigma_D}\frac{E}{(KL)^D}
\end{eqnarray}
Note that the dimensions of the density of states and of the level spacing are $[\nu]=(1/\mbox{volume})\times(1/\mbox{energy})$ and $[\delta(E)]=\mbox{energy}$.

 In what follows we will also need the density of states and the level spacing with respect only to fully symmetric states. They coincide with $\delta_1(E)$ and $\nu_1(E)$, no matter what the real $D$ is:
\begin{eqnarray}
\nu_1(E)\approx \frac{1}{\pi \alpha}\frac{K}{E},\quad \delta_1(E)=[L\nu_1(E)]^{-1}=\pi\alpha\frac{E}{KL}.
\label{levsp}
\end{eqnarray}
Note that for small $E\ll 1$ the density of states  becomes very small and, therefore, the level spacing  becomes relatively large. 

\section{The localized states and the optimal fluctuations\label{optimal3}}

For small disorder $W\ll 1$ 
 there is  some (exponentially small) number of localized states with  $E<0$, associated with  exponentially rare local fluctuations of the random potential. 
Let us look at the contribution of these localized states to the density of states. Following the standard procedure~\cite{Lifshitz, LifshitzBook} of finding an optimal fluctuation $V_{\bf j}$ and the corresponding localized wave-function
 $\Psi_{\bf n}$ we should minimize the functional
 \begin{widetext}
\begin{eqnarray}
\tilde{S}(\{\Psi,V\},\lambda,\eta)=\frac{1}{2}\sum_{\bf j}V^2_{\bf j}-\lambda\left\{\sum_{\bf jj^\prime}\Psi^*_{\bf j^\prime}\varepsilon_{\bf j-j^\prime}\Psi_{\bf j}+\sum_{\bf j}V_{\bf j}|\Psi_{\bf j}|^2-E\right\}-\eta\left\{\sum_{\bf j}|\Psi_{\bf j}|^2-1\right\}
\label{h700}
\end{eqnarray}
\end{widetext}
with respect to two functions $\Psi_{\bf j}$, $V_{\bf j}$ and two additional parameters $\lambda$ and $\eta$. Variation of \eqref{h700} with respect to $V_{\bf j}$ allows one to express $V_{\bf j}$ through $\Psi_{\bf j}$ and $\lambda$:
\begin{eqnarray}
V_{\bf j}=\lambda|\Psi_{\bf j}|^2
\label{h701}
\end{eqnarray}
and we are left with the functional
\begin{eqnarray}
-\frac{1}{\lambda}\tilde{S}(\{\Psi \},\lambda)=\sum_{\bf jj^\prime}\Psi^*_{\bf j^\prime}\varepsilon_{\bf j-j^\prime}\Psi_{\bf j}+\frac{\lambda}{2}\sum_{\bf j}|\Psi_{\bf j}|^4-E\sum_{\bf j}|\Psi_{\bf j}|^2
\label{h702}
\end{eqnarray}
subject to minimization with respect to $\Psi_{\bf j}$ with the normalization constraint
\begin{eqnarray}
\sum_{\bf j}|\Psi_{\bf j}|^2=1
\label{h703}
\end{eqnarray}
Thus, we arrive at the nonlinear Schr\"{o}dinger equation
\begin{eqnarray}
\sum_{\bf j^\prime}\varepsilon_{\bf j-j^\prime}\Psi_{\bf j^\prime}+\{\lambda|\Psi_{\bf j}|^2-E\}\Psi_{\bf j}=0
\label{h704}
\end{eqnarray}

 The function $\Psi_{\bf j}$ should be localized, i.e., it should vanish for large $_{\bf j}$. The implications of this requirement we will discuss in the Section~\ref{Localized vs delocalized wave-functions}.
 
 Finally, we have to ensure that the normalization condition \eqref{h703} is fulfilled. To satisfy this condition we have to choose the only free parameter at our disposal -- $\lambda$.

The explicit form of the wave function $\Psi^{\rm (opt)}_{\bf j}$ and optimal parameter $\lambda_{\rm opt}$ can only be found by means of numerical solution of the essentially discrete nonlinear Schr\"{o}dinger equation \eqref{h704}. The final expression to the optimal exponent in \eqref{h1er} reads
\begin{eqnarray}
\frac{S_{\rm opt}}{W^2}=\frac{\sum_{\bf j}\left(V^{\rm (opt)}_{\bf j}\right)^2}{2W^2}=\frac{\lambda_{\rm opt}^2}{2W^2}\sum_{\bf j}\left|\Psi^{\rm (opt)}_{\bf j}\right|^4
\label{h7044}
\end{eqnarray}
We are interested in the behavior of $S_{\rm opt}(E)$ for small energies $|E|\ll 1$, so that $S_{\rm opt}(E)$ can be expanded in $E$ up to linear terms:
\begin{eqnarray}
S_{\rm opt}(E)\approx S_{\rm opt}(0)+\lambda_{\rm opt}(0)E.
\label{h7045}
\end{eqnarray}
Note that $S_{\rm opt}(0)$ and $\lambda_{\rm opt}(0)$ are some numerical constants of order unity, depending on $D$, $\alpha$ and on the type of lattice.

\section{The  local character of the optimal  fluctuation}

The equations \eqref{h702}, \eqref{h704} are perfectly standard -- they do not differ from what we have for the conventional Lifshits tails, arising in the case of $\alpha>D/2$. Then why do we expect an anomalous behavior of the tails in our case $\alpha<D/2$?

Let us model an optimal fluctuation as a square potential well with depth $U$ and width $a$, so that we have to minimize the function of two variables
\begin{eqnarray}
S(U,a)\sim U^2 a^d
\label{h704e}
\end{eqnarray}
To have a level with energy $E$ this well should obey the following constraints
\begin{enumerate}
\item The well should be deeper than $E$: ($U>|E|$)
\item The well should be wider than the wave-length: $a>Q^{-1}=U^{-1/\alpha}$.
\end{enumerate}
It seems plausible that the narrowest possible well  is a good choice. Then, assuming $a\sim a_{\min}=Q^{-1}$ we have to minimize the function 
\begin{eqnarray}
S(U)\sim U^{2-D/\alpha}
\label{h704eeeee}
\end{eqnarray}
If $\alpha>D/2$ (as it is for conventional Lifshits tails with $\alpha=2$ and $D<4$) then $S$ decreases with decreasing $U$, so that the optimal fluctuation corresponds to minimal possible $U_{\min}\sim |E|$ which leads to the standard Lifshits result:
\begin{eqnarray}
S^{\rm (opt)}_{\rm Lif}\propto |E|^{2-D/\alpha}
\label{h704eew}
\end{eqnarray}

In our case $\alpha<D/2$ and $S$ decreases with decreasing $U$, so the minimum of $S$ corresponds to the deepest possible fluctuation. Thus, within the continual approximation the optimal fluctuation would be infinitely deep and infinitely narrow. In reality, however, the fluctuation should contain at least one site, so the minimum is attained at $a\sim 1$, $U\sim\max\{|E|,t\}$. As a result, we obtain
\begin{eqnarray}
S^{\rm (opt)}_{\rm nonLif}\sim 1
\label{h704ek}
\end{eqnarray}

\subsection{The  Flat Band Approximation (FBA)}

For very small $\alpha\ll 1$ the electrons are almost dispersionless in the main part of the Brillouin Zone
\begin{eqnarray}
\varepsilon({\bf k})\approx W_{\rm band},\quad E_{\rm loc}^{(0)}\approx W_{\rm band},
\label{h701noa}
\end{eqnarray}
the dispersion is only present in the domain of exponentially small $k\sim e^{-1/\alpha}$. In the leading approximation both the optimal potential
\begin{eqnarray}
V_{\bf j}^{\rm (opt)}=(-W_{\rm band}+E)\delta_{\bf j,0},
\label{h701no}
\end{eqnarray}
and the corresponding wave-function
\begin{eqnarray}
\Psi_{\bf j}^{\rm (opt)}=\delta_{\bf j,0}
\label{h701no1}
\end{eqnarray}
are perfectly localized at the same site.
\begin{eqnarray}
S^{\rm (opt)}(E)=\frac12 (W_{\rm band}-E)^2
\label{h701no2}
\end{eqnarray}

\subsection{The  Single Site Approximation (SSA)}

If $\alpha$ is not specially small, the FBA does not work: the wave function is not localized at one site, so that formula \eqref{h701no1} is not valid. However, as we conclude from numerics (se Section~\ref{Numerical study of the optimal fluctuation}), the potential $V_{\bf j}^{\rm (opt)}$ remains extremely short range even for $\alpha$ away from zero: the potential remains localized at a single site with an accuracy better than 1\%!
Thus, it is very interesting to explore the single-site approximation (SSA) that postulates \begin{eqnarray}
V_{\bf j}^{\rm (opt)}=V_0(E)\delta_{\bf j,0},\quad S^{\rm (opt)}(E)=V_0^2(E)/2,
\label{h701n}
\end{eqnarray}
where the dependence $V_0(E)$ is yet to be found.
We stress again that, strictly speaking, the formula \eqref{h701n} is incorrect. Namely, it is inconsistent with the requirement \eqref{h701} which relates the shape of optimal potential to that of the optimal wave-function. Nevertheless, as it is demonstrated by the numerical results of the Section~\ref{Numerical study of the optimal fluctuation}, SSA works extremely  well, as long as we are interested in the ''integral''  characteristics, governed by the core of fluctuation. What is also important, SSA allows for the analytical solution of the arising quantum-mechanical problem. In particular, in \cite{TikhonovIoselevichFeigel'man} it was shown that, within SSA
\begin{eqnarray}
V_0(E)=\left\{\dashint_{BZ}\frac{d^D{\bf k}}{(2\pi)^D}\frac{1}{E-\varepsilon_{\bf k}}\right\}^{-1}\label{h13j}
\end{eqnarray}

However, we choose to postpone using the SSA, because there are many important and nice results that can be derived without appealing to any approximation.

\section{Localized vs delocalized wave-functions: general consideration\label{Localized vs delocalized wave-functions}}

As long as we consider systems of finite size, the optimal fluctuation method is perfectly applicable not only to genuine localized states with $E<0$, but to all the states, including those with $E>0$.

In the Section~\ref{optimal3} we have studied only the electronic ground state in the presence of the optimal fluctuation, here we will discuss the entire spectrum of the states. We will see that, besides the standard fully delocalized states with positive energies (plane waves),  there is a lot of hybrid states -- partly localized and partly delocalized.

Suppose that we have found the form of optimal fluctuation $V^{\rm (opt)}_{\bf j}$. 
To find the entire set of the states $\psi_{\bf j}^{(m)}$ and the corresponding energies $E_m$, we have to solve the linear Schr\"{o}dinger equation
\begin{eqnarray}
\sum_{\bf j^\prime}\varepsilon_{\bf j-j^\prime}\psi_{\bf j^\prime}^{(m)}+\{V^{\rm (opt)}_{\bf j}-E_m\}\psi^{(m)}_{\bf j}=0,
\label{h704ew}
\end{eqnarray}
to apply periodic boundary condition to wave-functions $\psi_{\bf j}^{(m)}$, and obtain a discrete set of eigenenergies $E_{m}$ and the corresponding eigenfunctions $\psi_{m}({\bf j})$. Clearly, $\Psi^{\rm (opt)}_{\bf j}$ will be one of these states (the ground state with energy $E_0$).
A formal solution of \eqref{h704ew} may be written as
\begin{eqnarray}
\psi_{\bf j}=\sum_{\bf j^\prime}g_{E}({\bf j-j^{\prime}})\psi_{\bf j^{\prime}}V_{\bf j^{\prime}}^{\rm (opt)},
\label{h704vg}
\end{eqnarray}
where
\begin{eqnarray}
g_{E}({\bf j,j^{\prime}})=g_{E}({\bf r})=\sum_{\bf n}\frac{\exp[i({\bf k_{\bf n}\cdot r})
]}{E-\varepsilon_{\bf n}},\quad {\bf r}\equiv {\bf j-j^{\prime}}.
\label{h8fg}
\end{eqnarray}
is the Green function of the free Schr\"{o}dinger equation. Note that there is no free term in the solution \eqref{h704vg}  since we have assumed that the energy $E$ is out of resonance with all the eigenfrequencies of the free Schr\"{o}dinger equation: $E\neq \varepsilon_n$ for all $n$. Writing $\psi_n({\bf j})$ in terms of the Green function \eqref{h8fg}, which uses the basis \eqref{h3}, ensures that
the  boundary conditions for the wave function are fulfilled automatically.

The sum over ${\bf j^\prime}$ in \eqref{h704vg} is dominated by small $|{\bf j^\prime}|\sim 1$, because we have assumed that $V_{\bf j^\prime}^{\rm (opt)}$ is localized: it rapidly decays with $|{\bf j^\prime}|$. Therefore, for ${\bf j}\gg 1$ we get
\begin{eqnarray}
\psi_{\bf j}=A(E)g_{E}({\bf j})
\label{h704vg1}
\end{eqnarray}
where $A(E)$ is certain ${\bf j}$-independent coefficient.
Thus, the asymptotics of the wave function feels the presence of the optimal fluctuation only through the value of the energy $E$. There are two different cases that we will discuss: negative energies $E < 0$ and positive energies $E>0$.

\subsection{Negative energies: localized wave-function}

When the energy of the state is negative, the Green function can be approximated by the integral instead of the discrete sum
\begin{eqnarray}
    g^{\rm (loc)}_{E} = \int_{\rm BZ}\frac{d^D {\bf k}}{(2\pi)^D}\frac{e^{i {\bf (k \cdot r)}}}{E - \varepsilon({\bf k})},
    \label{locState}
\end{eqnarray}
since it converges at $k$ in the entire Brillouin zone. The large-$r$ asymptotic behavior of $g_{E}^{\rm(loc)}({\bf r})$ can be easily evaluated. At smallest distances $|{\bf r}|\lesssim r_0\sim 1$ the components with high momenta $k\sim\pi$ give principal contribution to \eqref{locState}, $E$ in the denominator can be neglected compared to $\varepsilon({\bf k})$ and we get $g\sim 1$ in this range of distances. However,  ${E}$ in the denominator still can be neglected in a wider range, namely, for $r\lesssim r_1$ where
\begin{eqnarray}
r_1(E)\sim1/K\sim {E}^{-1/\alpha}\gg 1
\end{eqnarray}
In this range  of distances ($r_0\ll r\ll r_1$) we have:
\begin{eqnarray}
g_{E}^{\rm(loc)}({\bf r})\approx-\int_{BZ}\frac{d^D{\bf k}}{(2\pi)^D}\frac{e^{i ({\bf k\cdot r})
}}{\varepsilon({\bf k})}\approx-\frac{1}{r^{D-\alpha}}\int\frac{d^D{\bf q}}{(2\pi)^D}\frac{e^{i ({\bf q\cdot m})}}{\varepsilon(q)}\propto\frac{1}{r^{D-\alpha}},
\label{nooo2e}
\end{eqnarray}
where we have introduced ${\bf m}\equiv {\bf r}/r$ and ${\bf q}\equiv {\bf k}r$.

The main contribution to the integral \eqref{nooo2e} here comes from $q\sim 1$, or, from relatively small $k\sim 1/r$.

For $r\gg r_1(E)$ we can expand the integrand in $\varepsilon(k)$ and get
\begin{eqnarray}
g_{E}^{\rm(loc)}({\bf r})\approx\frac{1}{E^2}\int_{\rm BZ}\frac{d^D{\bf k}}{(2\pi)^D}e^{i ({\bf k \cdot r})}\varepsilon(k)=\frac{1}{E^2r^{D+\alpha}}\int\frac{d^D{\bf q}}{(2\pi)^D}e^{i ({\bf q \cdot m})}\varepsilon(q)\propto\frac{1}{r^{D+\alpha}},
\label{noo3}
\end{eqnarray}
It is easy to see that the results \eqref{nooo2e} and \eqref{noo3} match at $r\sim r_1$.

Thus
\begin{eqnarray}
g_{E}^{\rm(loc)}({\bf r})\sim \left\{\begin{aligned}
1,\quad & r\lesssim r_0,\\
r^{\alpha-D},\quad &r_0\ll r\ll r_1,\\
r_1^{2\alpha}(E)r^{-\alpha-D},\quad &r\gg r_1,
\end{aligned}
\right.
\label{noo4}
\end{eqnarray}
and
\begin{eqnarray}
\psi_{\rm opt}^{\rm(loc)}({\bf r})=\frac{1}{c}g_{E}^{\rm(loc)}({\bf r})
\label{noo4b}
\end{eqnarray}
where $c\sim 1$ is the normalization constant. Note that the main contribution to the normalization integral (and, therefore, to $c$) comes from the range $r\sim 1$, so that $c$ is almost $E$-independent. It should be mentioned that  the asymptotic formula \eqref{h704vg1} and, hence, the formula \eqref{noo4b} either, does not apply at $r\sim 1$. So, to evaluate $c$, one, in principle, has to use an explicit numerical solution of the initial discrete problem. In general,
the localized part is not strongly sensitive to $E$, so, for $E\ll 1$, 
\begin{eqnarray}
\Psi_{\rm opt}^{\rm(loc)}({\bf r})\approx\frac{1}{c}g_{E=0}^{\rm(loc)}({\bf r})
\label{noo4bu}
\end{eqnarray}

\subsection{Positive energies: quasi-localized wave function}

The vast majority of the  eigenstates $\psi^{(m)}_{\bf j}$ are not much affected by the presence of the optimal fluctuation, so that the corresponding eigenfunctions and eigenenergies are described by \eqref{h3} and \eqref{h4}
\begin{eqnarray}
\psi^{(m)}_{\bf j}\approx \phi_{\bf n}({\bf j}), \quad E_{n}\approx\varepsilon_{\bf n},
\label{h704a}
\end{eqnarray}
Although these states are almost insensitive to the presence of the optimal fluctuation, scattering at the typical weak fluctuations leads to the power-law localization of them \cite{nosov2019correlation, modak2020manybody}. The corresponding localization length $l_E$ is inversely proportional to the strength of the disorder $W$ and, therefore, is much larger than the radius of the optimal fluctuation. 

In this paper we mostly consider the effects arising from the interaction of a particle with strong local fluctuations. Therefore, we focus on the states that are sensitive to the potential \eqref{h701}, -- the states, fully symmetric with respect to rotations around the center of the optimal fluctuation. Note, that the local level spacing within this subset is $\delta_1(E)\propto L^{-1}$ (see \eqref{levsp}), which, for $D>1$, is much larger than the total level spacing $\delta_D$. 

Still, as we will see soon, even these fully symmetric states are strongly delocalized, except for a bunch of $\sim M(E_0)$ states in a narrow interval of energies $|E-E_0|\lesssim\Delta(E_0)$ around $E_0$, where the states can be effectively localized.

\begin{figure}
\includegraphics[width=0.8\textwidth]{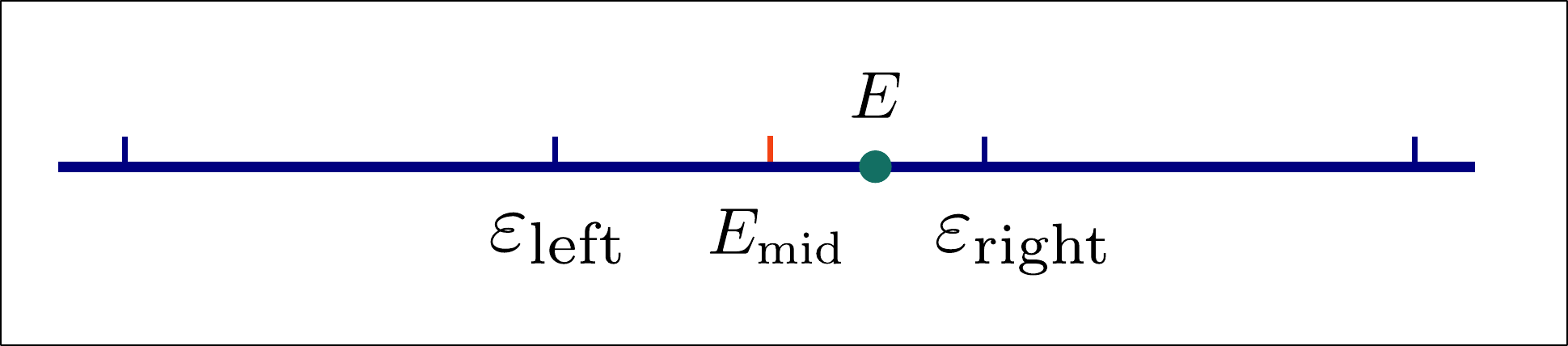}
\caption{Position of energy $E$ on the background of equidistant $\varepsilon_n$, and an illustration to the definition of quantities $\varepsilon_{\rm right}$, $\varepsilon_{\rm left}$, and $E_{\rm mid}$}
\label{levels2}
\end{figure}

Under which condition the wave-function with positive energy is effectively localized? To answer this question let us introduce an important characteristic
\begin{eqnarray}
\epsilon(E)\equiv\frac{E-\varepsilon_{\rm mid}}{\delta_1(E)},\quad \varepsilon_{\rm mid}(E)\equiv\frac{\varepsilon_{\rm right}+\varepsilon_{\rm left}}{2}
\end{eqnarray}
where $\varepsilon_{\rm left}$ is the closest neighbour of $E$ from the left, and $\varepsilon_{\rm right}$ -- from the right in the string of eigenenrgies $\varepsilon_n$, corresponding to free fully symmetric states (see Fig.~\ref{levels2}). The local level spacing is $\delta_1(E)=\varepsilon_{\rm right}-\varepsilon_{\rm left}$.

Suppose that the energy $E$ is placed in the middle of the interval $(\varepsilon_{\rm left},\varepsilon_{\rm right})$, or, in other words $E=E_{\rm mid}(E)$ and $\epsilon(E)=0$. Then, obviously, for $r \ll r_1$ the terms in the sum \eqref{h8fg} with $\varepsilon_n<E$ and with $\varepsilon_n>E$ will cancel each other in pairs exactly in a way, prescribed by the principal value integration. Hence, for $r \ll r_1$ and $\epsilon(E) = 0$ the Green function is given by the following integral
\begin{eqnarray}
    g_E^{(\epsilon = 0)}(r \ll r_1) = \dashint_{\rm BZ}\frac{d^D {\bf k}}{(2\pi)^D}\frac{e^{i {\bf (k \cdot r)}}}{E - \varepsilon({\bf k})},
    \label{QlocState1}
\end{eqnarray}
which is evaluated in exactly the same manner as before
\begin{eqnarray}
    g_{E}^{(\epsilon = 0)}({\bf r})\sim \left\{\begin{aligned}
1,\quad & r\lesssim r_0,\\
r^{\alpha-D},\quad &r_0\ll r\ll r_1
\end{aligned}
\right.
\label{QlocState2}
\end{eqnarray}

Evaluation of the very far tails $r \gg r_1$ is not so straightforward. Indeed, since $Kr \gg 1$ one needs to account for the discreteness of the system even when $\epsilon(E) = 0$. Explicitly, the Green function reads
\begin{eqnarray}
    g_E(|{\bf r}| \gg r_1) \propto \sum_{\bf n}\frac{e^{i {\bf k_n r}}}{E_{\rm mid} - \varepsilon_{\bf n}}.
    \label{QLocFarAway}
\end{eqnarray}
The main contirbution to this sum comes from $|{\bf k_n}| \approx K$, hence, we expand $\varepsilon_{\bf k_n}$ in the vicinity of $E$. Let us introduce integer $l$ in the following way
\begin{eqnarray}
    n=n_{\rm left}(E_{\rm mid})+l, \quad  k_n=K(E_{\rm mid})+\frac{\pi}{L}(l-1/2),\quad     \varepsilon_n=E_{\rm mid}+(l-1/2)\delta_1(E).
\end{eqnarray}
Since the spectrum is spherically symmetric, we need the asymptotic of the spherical wave
\begin{eqnarray}
    f_{n}(r)\approx x^{-(D-1)/2}
    \cos (x+\varphi_D), \quad x\equiv \left(K(E_m)r-(\pi r/L)[\epsilon-(l-1/2)]\right)\gg 1.
\end{eqnarray}
Therefore, relation \eqref{QLocFarAway} reads
\begin{equation}
    g_E(|{\bf r}| \gg r_1) \propto \text{Re} \left[ -e^{i K r - i\frac{\pi r}{2 L}}  (Kr)^{-(D-1)/2} \sum_{l=-\infty}^{\infty} \frac{e^{i\frac{\pi r}{L}l}}{l - \frac{1}{2}} \right],
    \label{QLocFarAway1}
\end{equation}
since it converges at small $l$'s. Now, we use
\begin{align}
    \sum_{l=-\infty}^{\infty} \frac{e^{i \pi l z}}{l-1/2} &= -i\pi e^{i \frac{\pi}{2}z},
\end{align}
and obtain the final expression for the wave function with positive energy in the middle of the interval $E = E_{\rm mid}$
\begin{eqnarray}
\Psi_E({\bf r})\sim \left\{\begin{aligned}
1,\quad & r\lesssim r_0,\\
r^{\alpha-D},\quad &r_0\ll r\ll r_1,\\
r_1^{\alpha - D}\frac{\sin{\left(K r + \varphi_D\right)}}{(K r)^{\frac{D-1}{2}}},\quad &r\gg r_1.
\end{aligned}
\right.
\label{noo411111}
\end{eqnarray}
The oscillating tail at $r \gg r_1$ prevents $\Psi_E({\bf r})$ from being truly localized: even when $\epsilon = 0$ the wave function has delocalized tails. We call this state ``quasi-localized".

\subsubsection{Effective localization condition}

We consider finite systems of size $(2L)^{D}$, hence, it is possible for the quasi-localized state to be effectively localized in the vicinity of the optimal fluctuation. Indeed, one can compute the norm of $\Psi_E({\bf r})$
\begin{widetext}
\begin{eqnarray}
    \int d^D {\bf r} |\Psi({\bf r})|^2 \sim \left(1 + r^{2(\alpha - D)}_1 K^{-D}\int_{1}^{K r}  d y \sin^2 y \right)
    \sim \left( 1 + E^{-\frac{2}{\alpha}(\alpha - D) - \frac{D}{\alpha} + \frac{1}{\alpha}}  L \right)\sim \left(1 + E^{-2 + \frac{D+1}{\alpha}} L \right).
\end{eqnarray}
\end{widetext}
The contribution from the oscillating tail vanishes when the energy is sufficiently low. Let us introduce $L_c(E)$ in the following way
\begin{eqnarray}
    L_c \equiv E^{-\frac{D+1-2\alpha}{\alpha}}.
    \label{Lc1}
\end{eqnarray}
Tail contribution vanishes if
\begin{eqnarray}
    L \ll L_c.
    \label{LocCond2}
\end{eqnarray}
Our calculations are valid only if we consider higly excited state, i.e. $KL \gg 1$, as given by condition \eqref{highExc}. Conditions \eqref{highExc} and \eqref{LocCond2} can be satisfied simultaneously in very large systems only if $\alpha < D/2$.

\begin{figure}[ht]
\includegraphics[width=\textwidth]{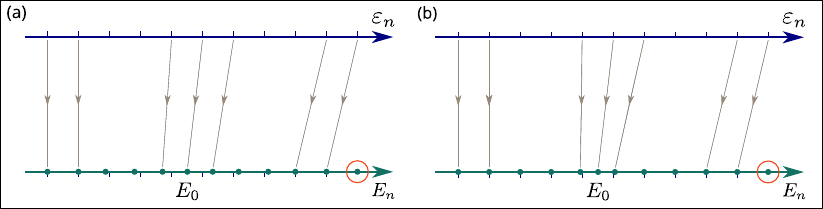}
\caption{Upper string: a locally equidistant spectrum of levels $E_n\equiv\varepsilon_n$ in the absence of the fluctuation. Lower string: a spectrum in the presence of fluctuation with (a) $L \ll L_c$, (b) $L \gg L_c$.. The levels $E_n$ within the localization energy domain are shifted with respect to $\varepsilon_n$. In general, they contain both localized and delocalized components.}
\label{levels1}
\end{figure}

Quasi-localized states, that we have just introduced, exist due to the presence of strong local fluctuations which correspond to the saddle-point solution. Since typical fluctuations are always present in real systems, one needs to take them into account. Let us demonstrate for the most simple case $D=1$ that the quasi-localized states are robust to these fluctuations. As we will show later (see Section~\ref{eigSpectrum}), the level spacing is not very sensitive to the presence of one potential fluctuation, hence, we can assume it to coincide with the one in clean system. Using that, we can easily find energy that corresponds to the level spacing $\delta_D(E)$ which is of order of the characteristic scale of the matrix element of the random potential $\sqrt{\langle V^2\rangle} \sim  W L^{-1/2}$:
\begin{eqnarray}
    E^\prime_c \sim W^{-\frac{\alpha}{1 - \alpha}} L^{-\frac{\alpha}{2(1-\alpha)}}.
\end{eqnarray}

Therefore, states with energies $E \ll E^\prime_c$ remain almost unperturbed owing to typical fluctuations. Some of them extend over the whole system: the localization length $l_E \sim W^{-2}E^{2 - \frac{2}{\alpha}}$~\cite{TikhonovIoselevichFeigel'man} for these energies is much larger than the system size (we call such states ``quasi-extended"); and some of them are quasi-localized in the sense described above, since $E_c \ll E^\prime_c$, where

\begin{eqnarray}
    E_c \sim L^{-\frac{\alpha}{D+1-2\alpha}}.
\end{eqnarray}

\section{Numerical study of the optimal fluctuation\label{Numerical study of the optimal fluctuation}}

We perform a numerical study of the optimal fluctuation dropping contribution from the delocalized tails. Indeed, because we know that any state with positive energy is either quasi-extended or quasi-localized, one cannot fine-tune the energy to remove the oscillating non-decaying contribution.

\begin{figure}[h!]
\includegraphics[width=0.6\textwidth]{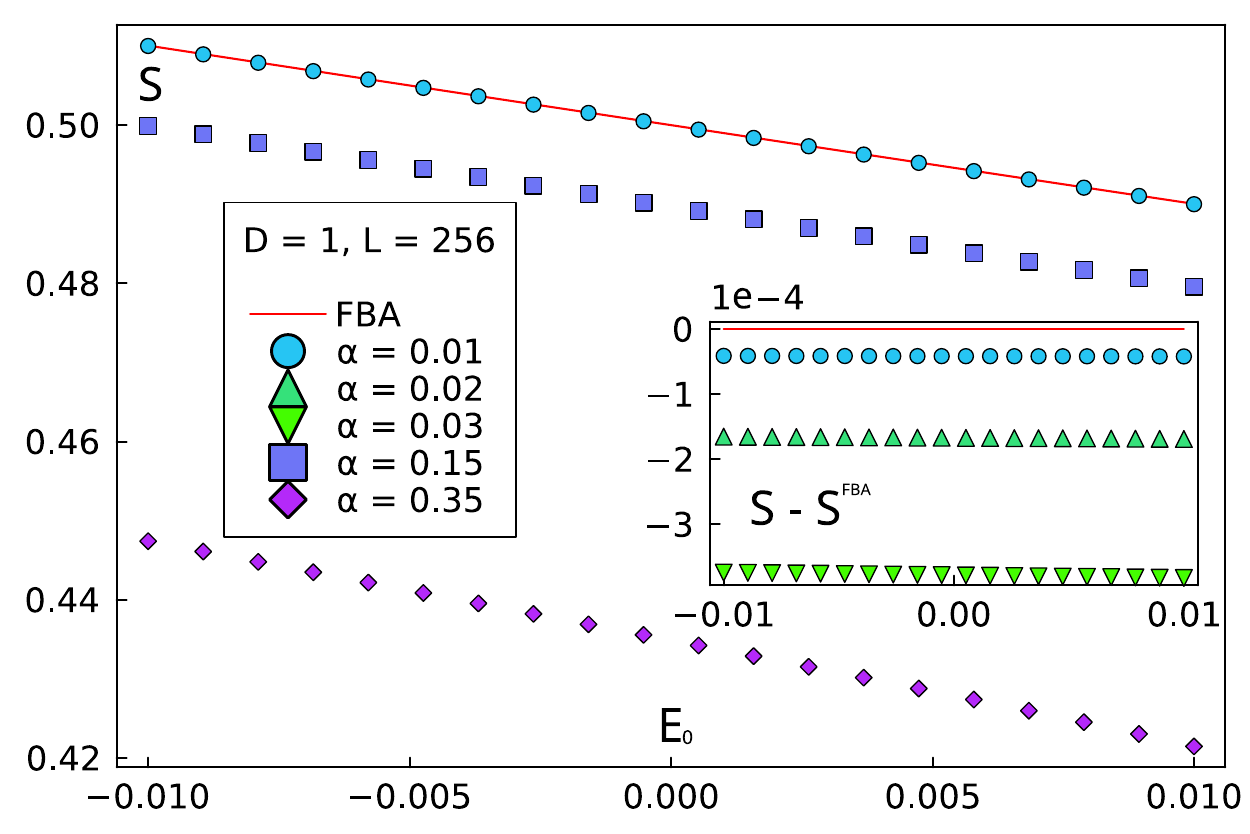}
\caption{The dependence $S^{\rm (opt)}(E_0)$ for $D=1$ obtained numerically. Solid line (red online) shows the result of FBA.}
\label{action1}
\end{figure}

\begin{figure}[ht]
\includegraphics[width=\textwidth]{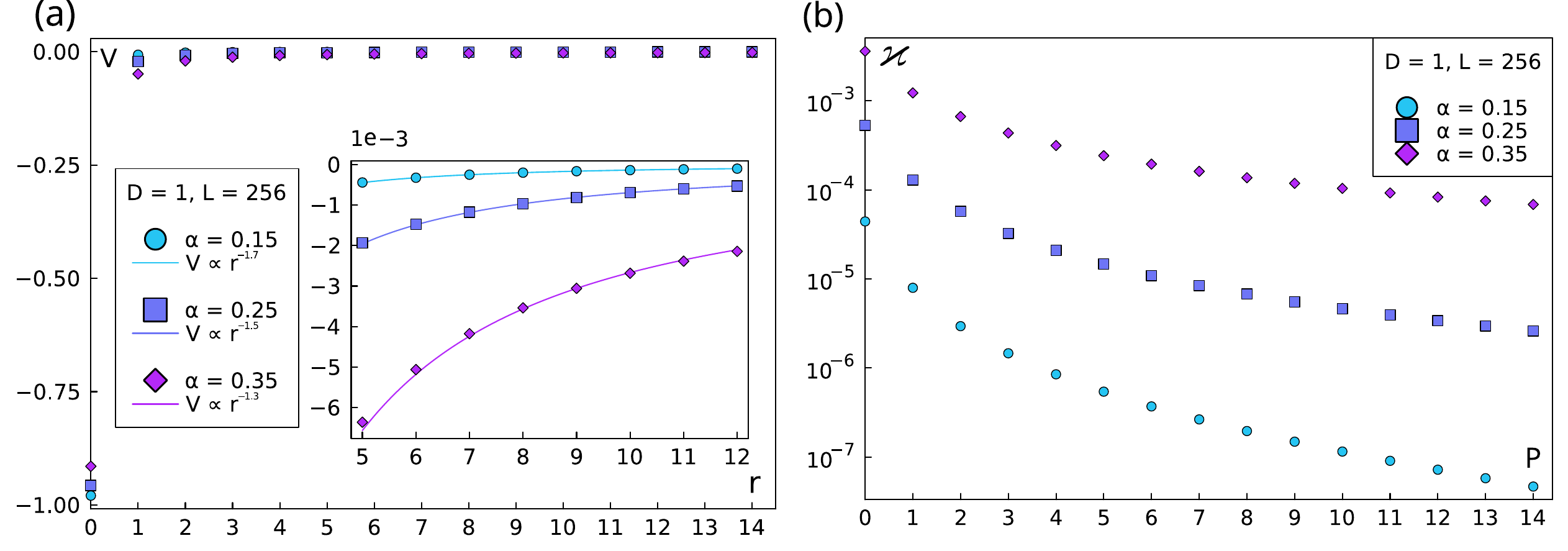}
\caption{(a) The shape of optimal fluctuation in $1D$. At first coordination sphere it drops already by two orders of magnitude, while in the tail it decreases only slowly (see inset). (b) The relative accuracy   $\varkappa(P) = [S^{(P)} - S^{(\infty)}]/ S^{(\infty)}$ of truncated models with $P$ shells of nonzero potentials.}
\label{potential1}
\end{figure}

Our results support strongly localized character of the core of the optimal fluctuation. At small $|{\bf j}|$ the potential $V_{\bf j}^{\rm (opt)}$ rapidly decays with $|{\bf j}|$. For example, in $1D$-case, $V_{\pm 1}^{\rm (opt)}/V_{0}^{\rm (opt)}$ varies from $0.01$ at $\alpha=0.15$ to $0.04$ at $\alpha=0.35$ (see Fig.~\ref{potential1}).

At the same time at $|{\bf j}|\gg 1$ the decay of $V_{\bf j}^{\rm (opt)}$ becomes rather slow and is well described by a power law:
\begin{eqnarray}
V_{\bf r}^{\rm (opt)}\propto |\Psi_{\bf r}^{\rm (opt)}|^2\propto |{\bf r}|^{2\alpha-2D}
\end{eqnarray}
which is perfectly consistent with the exact relations \eqref{noo4} and \eqref{h701}. Although the validity of the latter relation signals about the validity of our numerics, it shold be admitted that for the vast majority of questions which we address in this study, the tails of the potential $V_{\bf j}^{\rm (opt)}$ are irrelevant.

\begin{figure*}
\includegraphics[width=\textwidth]{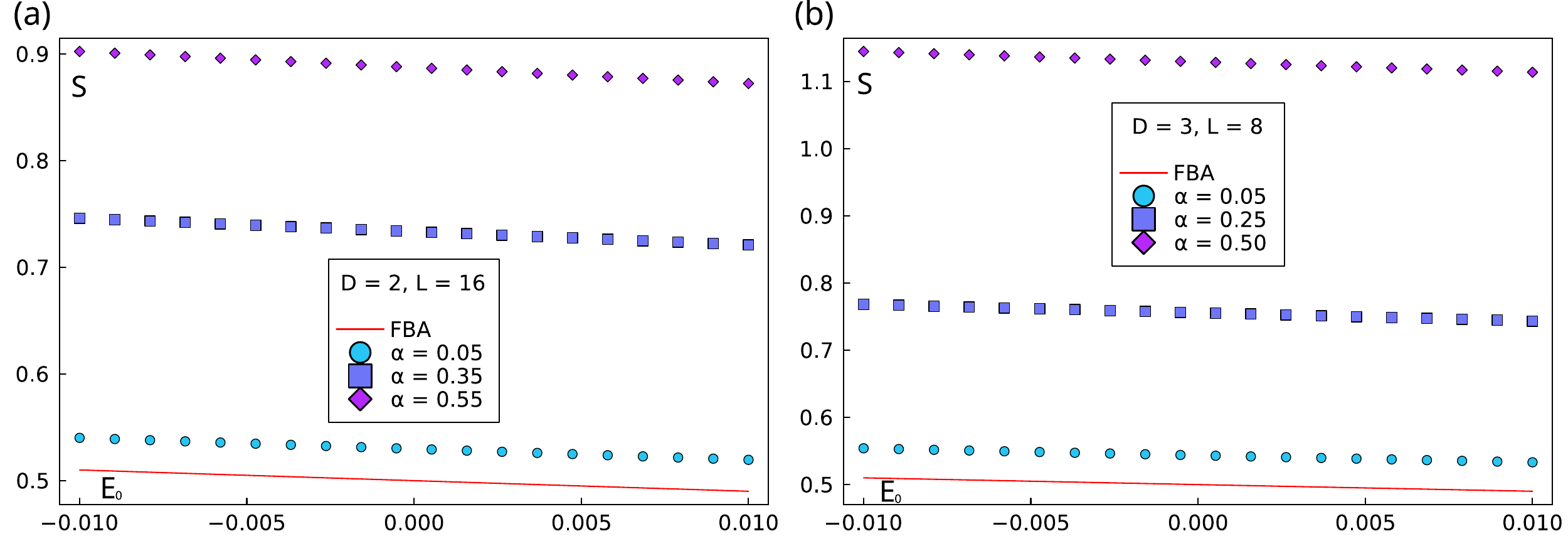}
\caption{The dependence $S^{\rm (opt)}(E_0)$ for (a) $D=2$, (b) $D=3$. Solid line (red online) shows the result of FBA.}
\label{action23}
\end{figure*}

To test the accuracy of the SSP we have found $S_{\rm opt}^{(P)}$   for series of truncated models where all  $V_{\bf j}^{\rm (opt)}$ were forcefully set to be zeroes for $|j|>P$, while the remaining $2DP+1$ potentials were chosen to optimize $S$. Particularly, due to \eqref{h701} we set

\begin{eqnarray}
    V^{\rm (opt)}_{|{\bf j}| \le P} = \lambda |\psi_{\bf j}|^2,\quad V^{\rm (opt)}_{|{\bf j}| > P} = 0.
\end{eqnarray}
\\
After that, we add normalization condition $\sum_j |\psi_j|^2 = 1$ and solve the system of $P+2$ equations (instead of $2DP+1$ since the localized state possesses discrete rotational symmetry). During the calculations, $g^{\rm (loc)}_{E}$ is used instead of $g_{E}$ since we are interested in the localized solution. The results of exact optimization are illustrated in Fig.~\ref{action1} and Fig.~\ref{action23}.

\section{Away from $E_0$:
 Partly localized wave-functions\label{spectrumm}}

In Section~\ref{Localized vs delocalized wave-functions} we have found the condition for the state of energy $E = E_{\rm mid}$ to be effectively localized and have studied the properties of the quasi-localized wave functions in detail. Now we discuss the properties of the states with energies $E_m\neq E_0$. Slightly away from $E_0$ we expect the $\epsilon \neq 0$ part to be small:
\begin{eqnarray}
\psi_{E_m}^{\rm(m,del)}({\bf r})\propto g_{E_m}^{(\epsilon \neq 0)}({\bf r})\propto [E_m-E_{\rm mid}(E_m)].
\label{noo1rjj}
\end{eqnarray}
and, since $E - E_{\rm mid}(E_m)=0$ at $E_m=E_0$, at small $E_m-E_0$ we will have have $E - E_{\rm mid}(E_m)\propto E_m-E_0$.

\begin{figure}[ht]
\includegraphics[width=0.8\textwidth]{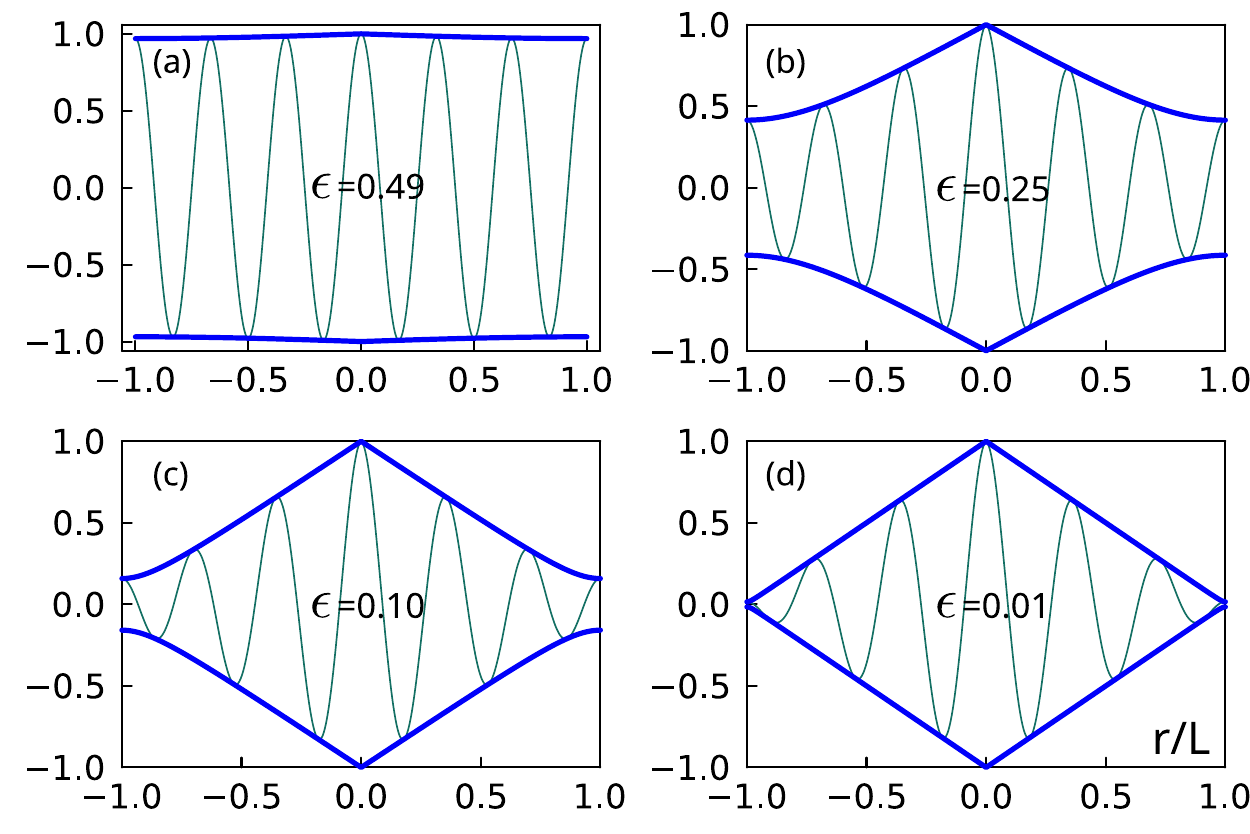}
\caption{The evolution of spatial shape of the delocalized part of the wave function in $D=1$ with the change of parameter $\epsilon$.  (a):  $\epsilon$=0.49, (b): $\epsilon$=0.25, (c): $\epsilon$=0.1, (d): $\epsilon$=0.01}
\label{deloc}
\end{figure}

\begin{widetext}
\subsection{The delocalized part of  the wave-function}

Let us again introduce integer $l$, such that 
\begin{eqnarray}
n=n_{\rm left}(E_m)+l, \quad \varepsilon_n=E_{\rm mid}(E_m)+(l-1/2)\delta_1, \quad k_n=K(E_m)-\frac{\pi}{L}[\epsilon-(l-1/2)],\\
E_m-\varepsilon_n=(E_m-E_{\rm mid}(E_m))-(l-1/2)\delta_1,\quad E_{\rm mid}(E_m)-\varepsilon_n=-(l-1/2)\delta_1,\\
f_{n}(r)\approx x^{-(D-1)/2}
\cos (x+\varphi_D),\quad x\equiv \left(K(E_m)r-(\pi r/L)[\epsilon-(l-1/2)]\right)\gg 1,
\end{eqnarray}
Then 
for the asymptotics of the $\epsilon \neq 0$ part of the Green function we can write
\begin{multline}
g_{E_m} ^{(\epsilon \neq 0)}({\bf r})\approx \sum_{l=-\infty}^{\infty}\phi_{n_{\rm left}(E_m)+l}(r)\phi^*_{n_{\rm left}(E_m)+l}(0)\left\{\frac{1}{[E_m-E_{\rm mid}(E_m)]-\delta_1(l-1/2)}-\frac{1}{-\delta_1(l-1/2)}\right\}\approx \\
\approx \frac{2K^{D-1}}{\sigma_D L}f(0)(Kr)^{-(D-1)/2}\sum_{l=-\infty}^{\infty}\left\{\frac{1}{[E_m-E_{\rm mid}(E_m)]-\delta_1(l-1/2)}-\frac{1}{-\delta_1(l-1/2)}\right\}\times\\\times \cos\{(Kr-(\pi r/L)[\epsilon-(l-1/2)])+\varphi_D\}
\approx\\\approx
\frac{2K^{D-1}}{\sigma_D L}f(0)(Kr)^{-(D-1)/2}\frac{1}{\delta_1}{\rm Re\;}\left\{\sum_{l=-\infty}^{\infty}\frac{\epsilon\exp\{iKr-i(\pi r/L)[\epsilon-(l-1/2)]+i\varphi_D\}}{[\epsilon-(l-1/2)](l-1/2)}\right\}=\\=
\frac{2K^{D-1}}{\sigma_D L}f(0)(Kr)^{-(D-1)/2}\frac{1}{\delta_1}{\rm Re\;}\left\{\exp(iKr+i\varphi_D) \sum_{l=-\infty}^{\infty}\frac{\epsilon\exp\{-i(\pi r/L)[\epsilon-(l-1/2)]\}}{[\epsilon-(l-1/2)](l-1/2)}\right\}=\\=
\frac{2f(0)}{\sigma_D}K^{D-1}\nu_1(E)(Kr)^{-(D-1)/2}{\rm Re\;}\left\{\exp(iKr+i\varphi_D) \Phi(r/L,\epsilon)\right\}
\label{fin-deloc1}
\end{multline}
where
\begin{eqnarray}
\Phi(z,\epsilon)=\sum_{l=-\infty}^{\infty}\frac{\epsilon e^{-i\pi z[\epsilon-(l-1/2)]}}{(\epsilon-(l-1/2))(l-1/2)}\approx\left\{\begin{aligned}-\pi\tan(\pi\epsilon)& \quad\mbox{for}\;z\ll 1,\;\mbox{any}\;\epsilon,\\
-\pi^2(1-|z|)\epsilon & \quad\mbox{for}\;\epsilon\ll 1,\;\mbox{any}\;z,\\
1/(\epsilon\mp 1/2)& \quad\mbox{for}\;\epsilon\to\pm1/2,\;\mbox{any}\;z,
\end{aligned}
\right.
\label{asymphi}
\end{eqnarray}
The corresponding contribution to the wave function $\psi_{\rm deloc}({\bf r})\propto g^{(\epsilon \neq 0)}({\bf r})$ is delocalized. Having in mind that  the preexponential coefficient in \eqref{fin-deloc1} is $L$-independent, we conclude that the normalization integral $N_{\rm deloc}=\int|\psi_{\rm deloc}(r)|^2r^{D-1}dr\propto L$. Thus, in the case of general $\epsilon\sim 1$, when also $\Phi(z,\epsilon)\sim 1$, the norm $N_{\rm deloc}\sim L$  strongly dominates over the norm of the localized part $N_{\rm loc}\sim 1$.

Since we are interested in such wave functions, that are at least partly localized (i.e., $N_{\rm loc}\gtrsim N_{\rm deloc}$), we have to concentrate on the case $\epsilon\ll 1$, when $\Phi(z,\epsilon)\ll 1$. Therefore we  are allowed to use the corresponding asymptotics of \eqref{asymphi}. As a result
\begin{eqnarray}
 g_{E} ^{(\epsilon \neq 0)}({\bf r})\approx 
-\frac{2\pi^2f(0)}{\sigma_D}K^{D-1}\nu_1(E)(Kr)^{-(D-1)/2}\epsilon(1-r/L)\cos(Kr+\varphi_D)=C\frac{\epsilon\sqrt{K^{D+1}L}}{E}\tilde{\phi}_{\rm deloc}(r),\label{fin-deloc2}\\ 
C=-\frac{\pi f(0)}{\alpha}\sqrt{\frac{2}{3\sigma_D}} = - \frac{2^{\frac{2-D}{2}}\pi}{\alpha \Gamma{(D/2)}}\sqrt{\frac{\pi^{\frac{2-D}{2}}\Gamma{(D/2 + 1)}}{3D}}
\label{fin-deloc2a}
\end{eqnarray}
where the normalized delocalized wave function $\tilde{\phi}_{\rm deloc}(r)$ has, for $|\epsilon|\ll 1$,
the following asymptotics at $Kr\gg 1$:
\begin{eqnarray}
\tilde{\phi}_{\rm deloc}(r)\approx\sqrt{\frac{6K^{D-1}}{\sigma_D L}}(Kr)^{-\frac{D-1}{2}}
(1-r/L)\cos (Kr+\varphi_D), 
\end{eqnarray} 
From \eqref{fin-deloc2} it is clear that the wave function becomes essentially delocalized already at 
\begin{eqnarray}
 \epsilon\gtrsim \sqrt{L_c(E)/L},\quad \mbox{where}\; L_c(E)\sim E^{2-2(D+1)/\alpha}\gg 1.\label{fin-deloc3}
\end{eqnarray}
\end{widetext}
When $\epsilon$ further increases and, finally, reaches $|\epsilon|\sim 1$ the shape of the localized wave function starts to change and gradually approaches the standard cosine form (see Fig.~\ref{deloc}).

It is now necessary to find the proper expression of $\epsilon$ as a function of the energy $E$. 

\section{Eigenenergies \label{eigSpectrum}}

Until now we didn't need the exact form of the optimal fluctuation and considered it to be short-range only. It is impossible to find the spectrum in the presence of the optimal fluctuation given by the solution of the nonlinear Shr\"odinger equation \eqref{h704} analytically. Hence, it is now when we use SSA explicitly. 

\subsection{The Dyson equation and its general solution}

The Dyson equation for the Green function $G_E({\bf r,r^{\prime}})$ reads
\begin{eqnarray}
G_E({\bf r,r^{\prime}})=g_E({\bf r,r^{\prime}})+Vg_E({\bf r,0})G_E({\bf 0,r^{\prime}})\label{h7}
\end{eqnarray}
Then, for $G_E({\bf r,r^\prime})$ we obtain
\begin{equation}
G_E({\bf r,r^{\prime}})=g_E({\bf r,r^{\prime}})+\frac{Vg_E({\bf r,0})g_E({\bf 0, r^{\prime}})}{1-g_E({\bf 0,0})V}=g_E({\bf r-r^{\prime},0})+\frac{Vg_E({\bf r,0})g_E({\bf 0, r^{\prime}})}{1-g_E({\bf 0,0})V}.
\label{h10}
\end{equation}
The eigenenergies $E_m$ of the corresponding Schr\"{o}dinger equation can be found as solutions of equations 
\begin{eqnarray}
g_E^{-1}({\bf 0,0})-V=0,\label{h11}
\end{eqnarray}
with respect to $E$. 
As earlier, we split the Green function into two terms
\begin{eqnarray}
g_E(0)=g^{(\epsilon = 0)}_E(0)+g^{(\epsilon \neq 0)}_E(0),\\
g^{(\epsilon = 0)}_E(0)= \dashint_{BZ}\frac{d^D{\bf k}}{(2\pi)^D}\frac{1}{E-\varepsilon({\bf k})}.\label{and7q1}
\end{eqnarray}
From the previous chapter, we know that when $E = E_{\rm mid}$ discrete part of the Green function is zero: $g^{\rm (deloc)}_E({\bf 0}) = 0$. Hence, we write
\begin{equation}
    g^{(\epsilon = 0)}_E(0)= \dashint_{BZ}\frac{d^D{\bf k}}{(2\pi)^D}\frac{1}{E-\varepsilon({\bf k})} \approx \sum_{n}\frac{|\psi_{n}(0)|^2}{E_{\rm mid}(E)-\varepsilon_{n}}.
\end{equation}
In order to evaluate singular part, $g^{(\epsilon \neq 0)}_E({\bf 0})$, we, again, introduce integer $l$
\begin{widetext}    
\begin{eqnarray}
    n=n_{\rm left}(E)+l, \quad \varepsilon_n=E_{\rm mid}(E)+(l-1/2)\delta_1\\
    E-\varepsilon_n=(E-E_{\rm mid}(E))-(l-1/2)\delta_1,\quad E_{\rm mid}(E)-\varepsilon_n=-(l-1/2)\delta_1, \\
    \delta_1 \equiv \delta_1(E).
\end{eqnarray}
Therefore, we find
\begin{multline}
    g^{(\epsilon \neq 0)}_E(0) \approx \frac{2K^{D-1} f(0)^2}{\sigma_D L}\sum_{l=-\infty}^{\infty}\left\{\frac{1}{[E-E_{\rm mid}(E)]-\delta_1(l-1/2)}-\frac{1}{-\delta_1(l-1/2)}\right\} = \\
    = -\frac{2K^{D-1}f(0)^2}{\sigma_D L}\frac{1}{\delta_1(E)}\sum_{l=-\infty}^{\infty}\frac{\epsilon}{[(1/2+l)-\epsilon][1/2+l]} =-\frac{2K^{D-1}f(0)^2}{\sigma_D}\nu_1(E)\pi \tan{(\pi \epsilon)}=-\pi \nu_D(E)\tan{(\pi \epsilon)}
    \label{G0deloc}
\end{multline}
\end{widetext}
Finally, we obtain
\begin{eqnarray}
 g^{\rm (deloc)}_E(0)=-\pi \nu_D(E)\tan{(\pi \epsilon)},
\end{eqnarray}
It is interesting that $D$-dimensional DOS $\nu_D(E)$ enters the final result.

Now, eigenenergies can be found from the following equation
\begin{eqnarray}
1/V=F_0(E)-\pi\nu_D(E)\tan(\pi\epsilon)
\end{eqnarray}
or
\begin{eqnarray}
\epsilon=\frac{1}{\pi}\arctan\left(\frac{F_0(E)-1/V}{\pi\nu_D(E)}\right)\label{rella},
\end{eqnarray}
where $F_0(E) = g^{(\epsilon = 0)}_E(0)$. Let us denote solution of
\begin{eqnarray}
F_0(E)-1/V=0
\end{eqnarray}
as $E=E_0(V)$. Hence, when energy $E$ is very close $E_0$: $|E-E_0(V)|\ll 1$ we find
\begin{equation}
\epsilon=\frac{1}{\pi}\arctan\left(\frac{E-E_0(V)}{\pi\nu_D(E)}\left.\frac{dF_0}{dE}\right|_{E=E_0(V)}\right)
=\frac{1}{\pi}\arctan\left(\frac{E-E_0(V)}{\Delta(E)}\right),
\label{expan1}
\end{equation}
where
\begin{align}
\Delta(E_0)=\frac{\pi\nu_D(E_0)}{b(E_0)}\sim \frac{K^D}{E}\ll E,\quad\mbox{(since  $\alpha<D/2$)},
\end{align}
and
\begin{align}
b(E_0)=\left.\frac{d F_0}{d E}\right|_{E=E_0(V)}\sim 1. 
\end{align}
When $E_0\ll 1$ we get
\begin{eqnarray}
b(E_0)\approx b(0)=-\int_{-\pi}^{\pi}\frac{d^Dk}{(2\pi)^D}\frac{1}{\varepsilon(k)^2}.
\end{eqnarray}
This integral safely converges $k\to 0$, since $\alpha<D/2$, and $b(E_0) \sim 1$ (see App.~\ref{b(E)}). 

Finally, we are in position to provide explicit expression for the eigenenergies spectrum. Every interval $(\varepsilon_n,\varepsilon_{n+1})$ contains only one energy level $E_n$
\begin{widetext}
\begin{eqnarray}
E_n=E_{\rm mid}(E_n)+\epsilon\delta_{1}=E_{\rm mid}(E_n)+\frac{\delta_{1}(E_0)}{\pi}\arctan\left(\frac{E_{\rm mid}(E_n)-E_0}{\Delta(E_0)}\right)
\approx\nonumber\\\approx\left\{\begin{aligned}
\varepsilon_{n}+\frac{\delta_{1}(E_0)\Delta(E_0)}{\pi(E_0-E_{\rm mid}(E_n))},\qquad & E_{\rm mid}(E_n)<E_0,\quad |E_{\rm mid}(E_n)-E_0|\gg\Delta(E_0)\\
E_{\rm mid}(E_n)+\delta_1(E_0)\left(\frac{E_{\rm mid}(E_n)-E_0}{\pi\Delta(E_0)}\right),\qquad &|E_{\rm mid}(E_n)-E_0|\ll\Delta(E_0),\\
\varepsilon_{n+1}-\frac{\delta_{1}(E_0)\Delta(E_0)}{\pi(E_{\rm mid}(E_n)-E_0)},\qquad & E_{\rm mid}(E_n)>E_0,\quad |E_{\rm mid}(E_n)-E_0|\gg\Delta(E_0)
\end{aligned}
\right. 
\end{eqnarray}
\end{widetext}
When $|E_n-E_0|\gg\Delta(E_0)$  energy level $E_n$ almost coincides with $\varepsilon_n$ or $\varepsilon_{n+1}$ and the corresponding wave function is almost unperturbed quasi-extended wave. When $|E_n-E_0|\ll\Delta(E_0)$ energy is very close to the middle of the interval $E_n\approx E_{\rm mid}(E_n)$, which corresponds to the quasi-localized state.

\subsection{Full expression for the wave function}

Let us now get back to the wave function. Since we are interested in the quasi-localized states with energies close to the $E_{\rm mid}$, we can expand relation \eqref{expan1} and plug it in the expression for the wave function. Hence, we obtain
\begin{equation}
    \psi_E(r) = \left[ 1  + u_1^2 L + u^2 L\right]^{-1/2} \left( \widetilde{\Psi}_{E_0} (r) + u_1 \sqrt{L} \psi^{\perp}_{n(E)}(r) + u \sqrt{L} \tilde{\phi}_{\rm deloc}(r) \right),
    \label{WFFinal}
\end{equation}
\begin{eqnarray}
    u_1 = \sqrt{\frac{\sigma_D}{2 L_c}}, \quad u = C^\prime E^{\frac{1-D}{2\alpha}}(E - E_0) \\
    C^\prime = -\frac{b(E_0) 2^{\frac{D+2}{2}} \Gamma^{\frac{3}{2}}\left(\frac{D}{2} + 1\right)}{3^{\frac{1}{2}} D^{\frac{3}{2}} \pi^{\frac{D+2}{2}}}
\end{eqnarray}
where $\widetilde{\Psi}_{E_0}(r) \sim r^{\alpha - D}$ -- localized part of the quasi-localized wave function and $\psi^{\perp}_{n(E)}(r)$ -- its delocalized tail at  $r > r_1$ that exists even for $E = E_0$:
\begin{eqnarray}
    \psi^{\perp}_{n(E)}(r) = r_1^{\alpha - D}\sqrt{\frac{2 L_c}{L \sigma_D}}\frac{\sin{\left(Kr + \varphi_D\right)}}{\left(K r\right)^{\frac{D-1}{2}}}.
\end{eqnarray}
Each of the functions $\widetilde{\Psi},\psi^{\perp},\tilde{\phi}_{\rm deloc}$ are normalized to unity.  We have also used the fact that three functions are orthogonal to each other (the overlap tends to zero as $1/L$).

The first two contributions to the overall normalization coefficient $\left[ 1 + u_1^2 L + u^2 L\right]^{-1/2}$ come from the quasilocalized part $\Psi_{E_0} (r)$, while the third  contribution arises due to deviation $E-E_0$.

Hence, when $|u|\sqrt{L} \ll 1$ and $L \ll L_c(E)$ states \eqref{WFFinal} are effectively localized. There is at least one such state with $E = E_0$ and $u = 0$. How many more of them are there? Effectively localized states should satisfy the following condition
\begin{eqnarray}
    |E - E_0| \ll \frac{E^{\frac{D-1}{2}}}{\sqrt{L}} \equiv \widetilde{\Delta}(E).
\end{eqnarray}
Therefore, there are $M_{\rm loc}$ more effectively localized states
\begin{eqnarray}
    M_{\rm loc} \equiv \frac{\widetilde{\Delta}(E)}{\delta_1(E)} \sim \frac{E^{\frac{D-1}{2\alpha}} L E^{\frac{1}{\alpha}}}{\sqrt{L} E} = \sqrt{\frac{L}{L_c}} \ll 1.
    \label{Mloc1}
\end{eqnarray}
Hence, there is only one effectively localized state in the vicinity of the optimal fluctuation.

\section{Inverse Participation Ratio}

In this Section  we will separately examine  cases $L\ll L_c$ \eqref{LocCond2}, and the opposite one $L \gg L_c$.

\subsection{IPR in the near tail: $L \ll L_c$}

In this case  $u_1\sqrt{L} \ll 1$ and one can neglect the second term in \eqref{WFFinal}. Then  for arbitrary $q$ and $D$ IPR obtains the following form
\begin{gather}
    P_q = \sum_{j} |\psi_E(j)|^{2q} \approx \frac{1 + u^{2q} L^{D(1-q) + q}}{(1 + u^2 L)^q}.
\end{gather}
If one fixes $q$, one immediately finds critical dimension
\begin{gather}
    D_{\text{cr}} = \frac{q}{q-1}.
    \label{Dcrit}
\end{gather}
If $D < D_{\text{cr}}$ it is possible to introduce two distinct characteristic lengths
\begin{eqnarray}
    \xi_1(E) \sim u^{-2}, \quad     \xi_2(E, q, D) \sim u^{-\frac{2q}{D(1-q) + q}}, \\ 1 \ll \xi_1(E) \ll \xi_2(E, q, D),
    \label{IPRxi}
\end{eqnarray}
Hence, IPR is given by the following relation
\begin{equation}
P_q(E, D)\approx\frac{1 + u^{2q} L^{D(1-q) + q}}{(1 + u^2 L)^q}\sim\left\{\begin{aligned}
1,\qquad & L\ll\xi_1,\\
\left( \frac{\xi_1}{L} \right)^{q},\qquad &\xi_1\ll L\ll \xi_2,\\
L^{-D(q-1)},\qquad & \xi_2\ll L\ll L_c.
\end{aligned}
\right.
\label{ipr3q}
\end{equation}

Case $D>D_{\rm cr}$ is much more surprising. Here $\xi_2(E)$ does not exist, IPR \eqref{ipr3q} is as follows
\begin{eqnarray}
P_q(E, D)\sim\left\{\begin{aligned}
1,\qquad & L\ll\xi_1,\\
\left( \frac{\xi_1}{L} \right)^{q},\qquad &\xi_1\ll L\ll L_c.
\end{aligned}
\right.
\label{ipr3qq}
\end{eqnarray}
For example, when $D=3$ and $q=2$ the IPR large-$L$ behavior is $P_2 \propto L^{-2}$ instead of the standard three-dimensional law $P_2 \propto L^{-3}$ even for energies far away from $E_0$, i.e. $\xi_1 \ll L$. 

If we define fractal dimension $D_q$ according to
\begin{eqnarray}
    P_q \sim L^{-D_q (q-1)}.
\end{eqnarray}
then, in our case, we obtain
\begin{eqnarray}
    D_q = \frac{q}{q-1} \quad \text{ when } \quad q > \frac{D}{D-1}
    \label{fracDim}
\end{eqnarray}
When $D>D_{\rm cr}$ it is easy to see from \eqref{fracDim} that the fractal dimension $D_q < D$.

\subsection{IPR in the far tail: $L \gg L_c$}

Here $u_1^2 L \gg 1$ and, therefore
\begin{eqnarray}
    P_q \approx \frac{1 + u_1^{2q} L^{D(1-q) + q} + u^{2q} L^{D(1-q) + q}}{(u_1^2 L + u^2 L)^q}.
\end{eqnarray}
In high dimensions, $D > D_{\rm cr}$, IPR, again, is fractal with the same fractal dimension
\begin{eqnarray}
P_q(E, D)\sim\left\{\begin{aligned}
\left( \frac{1}{u^2_1L} \right)^{q},\qquad & |E - E_0| \ll E^{\frac{D-\alpha}{\alpha}},\\
\left( \frac{1}{u^2 L} \right)^{q},\qquad & |E - E_0| \gg E^{\frac{D-\alpha}{\alpha}}.
\end{aligned}
\right.
\label{IPRLast}
\end{eqnarray}
Thus, we see that the localized part of the wave function can dominate IPR even when the state is not effectively localized (the norm is dominated by the delocalized tail).

\section{Conclusion}

We have demonstrated  that finite disordered systems with long range hopping indeed exhibit unusual properties. In addition to conventional localized states with negative energies that contribute to Lifshitz tails, the fluctuations of disorder in such systems support the existence of quasi-localized states with positive energies. The structure of such states is as follows: there is a strong short-range core, localized in the vicinity of a strong local fluctuation of disorder, and a weak  oscillating tail that spans  through the entire system. Under the condition $\eqref{LocCond2}$ contribution from the localized part of the wave function dominates the norm. However, as the systems size increases, the contribution of the tail increases either and sooner or later it overcomes the contribution of the core. It happens because  the long-range tails, however weak, decay too slowly and cannot be normalized  in an infinite system. Thus, the quasi-localized states can only exist in finite systems.

Note that the quasi-localized states can be highly excited states: there can be a lot of quasi-extended states with lower energies. 

\begin{figure}
\includegraphics[width=0.8\textwidth]{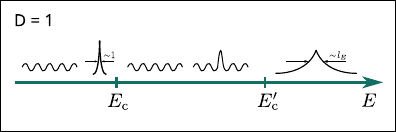}
\caption{Three different energy regimes exist in 1D systems: the lowest energies allow quasi-localized states to exist.}
\label{locdiagram}
\end{figure}

Moreover, even when condition \eqref{LocCond2} is not satisfied, and the norm of the wave function is dominated by the tail, the behavior of the IPR $P_q$ may still be determined by the localized core of the wave function. Then $P_q$ exhibits unusual behavior in a wide range of energies away from the energy of a quasi-localized state: for certain values of $q$ the character of $P_q$ is ``fractal''.

Found states are robust to typical fluctuations of the random potential. Keeping that in mind, in 1D, that is in the simplest possible case, we can distinguish three different energy domains as follows:
 \begin{itemize}
     \item $E \ll E_c$: here the quasi-localized states are formed on the continuum background of quasi-extended states.
    \item $E_c \ll E \ll E^\prime_c$: here remnants of the quasi-localized states become quasi-extended but exhibit unusual ``fractal'' properties.	
    \item $E \gg E^\prime_c$: here all the states are weakly localized owing to typical fluctuations, i.e. $l_E \ll L$.
\end{itemize}
See Fig.~\ref{locdiagram}.

\section*{Acknowledgements}

We are indebted to M.V.Feigel'man for valuable discussions, to I.M.Khaymovich for pointing out multiple helpful references, and to L.Levitov for useful comments on the manuscript. This work was supported by the Basic Research Program of The Higher School of Economics.

\bibliography{refs}

\clearpage

\appendix
\begin{widetext}
\section{Normalization function \label{b(E)}}

Throughout this paper, we very often encounter function that is defined as follows 
\begin{align}
    b(E) = \frac{\partial}{\partial E} \dashint_{\text{BZ}}\frac{d^D{\bf k}}{(2\pi)^D}\frac{1}{E - \varepsilon({\bf k})}
\end{align}
For $D=1$ it can be easily simplified
\begin{gather}
    b^{(D=1)}(E) = \dashint_{0}^{2^{\alpha}}\frac{\nu_0(E) - \nu_0(\varepsilon)}{(E - \varepsilon)^2}d\varepsilon - \nu_0(E) \frac{2^\alpha}{E(2^\alpha - E)} \sim 1\label{1b1}
\end{gather}
For small $E$ it is possible to treat \eqref{1b1} numerically. Plots of $b^{(D=1)}(E \ll 1)$ for various $\alpha$ can be found in Fig.~\ref{illB1}.

\begin{figure*}
    \includegraphics[width=\textwidth]{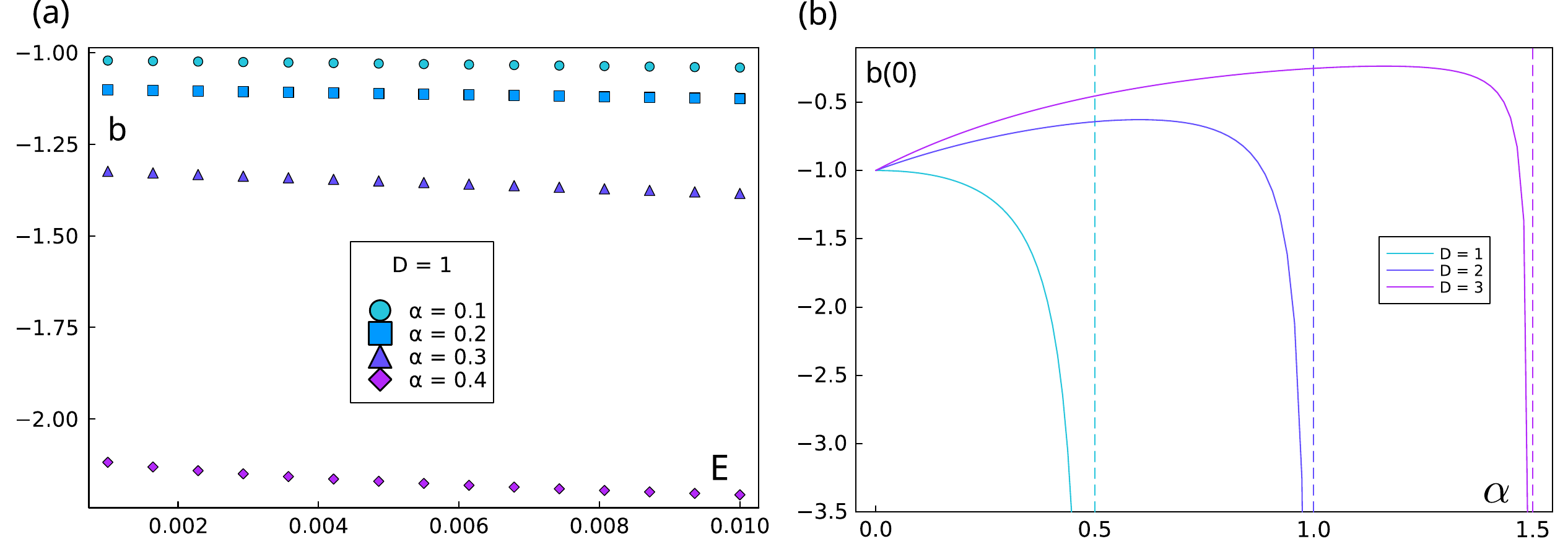}
    \caption{Left panel: numerical $b^{(D=1)}(E \ll 1)$ for $\alpha = 0.1, 0.2, 0.3, 0.4$ using eq.~\eqref{1b1}. Right panel: $b^{(D)}(0)$ for $D = 1, 2, 3$; vertical dotted lines show critical values $\alpha = D/2$.}
    \label{illB1}
\end{figure*}

In higher dimensions, $D>1$, it is impossible to obtain such a simple relation as we have for one dimensional case. Since we are interested in the behavior of $b(E)$ for very small energies we can can expand the denominator in small $|E| \ll 1$. Action of derivative leads to the following formula in the first non-zero order
\begin{eqnarray}
    b(E) \approx b(0)= -\int \frac{d^{D} {\bf k}}{(2\pi)^D} \varepsilon^{-2}(k).
\end{eqnarray}
$b(E)$ as a function of $\alpha$ is plotted in Fig.~\ref{illB1}.
\end{widetext}

\end{document}